\documentclass[11pt]{article}
\usepackage[pctex32]{graphics}

\def\ben{\begin{equation}}
\def\een{\end{equation}}

\def\nn{\nonumber} \def\bd{\begin{document}} \def\ed{\end{document}}
\def\ds{\documentstyle} \let\fr=\frac \let\bl=\bigl \let\br=\bigr
\let\Br=\Bigr \let\Bl=\Bigl
\let\bm=\bibitem
\let\na=\nabla
\let\pa=\partial \let\ov=\overline
\newcommand{\be}{\begin{equation}}
\newcommand{\ee}{\end{equation}}
\def\ba{\begin{array}}
\def\ea{\end{array}}
\def\ft#1#2{{\textstyle{\frac{\scriptstyle #1}{\scriptstyle #2} } }}
\def\fft#1#2{{\frac{#1}{#2}}}
\def\del{\partial}
\def\vp{\varphi}
\def\sst#1{{\scriptscriptstyle #1}}
\def\oneone{\rlap 1\mkern4mu{\rm l}}
\def\td{\tilde}
\def\wtd{\widetilde}
\def\ie{{\it i.e.\ }}
\def\dalemb#1#2{{\vbox{\hrule height .#2pt
        \hbox{\vrule width.#2pt height#1pt \kern#1pt
                \vrule width.#2pt}
        \hrule height.#2pt}}}
\def\square{\mathord{\dalemb{6.8}{7}\hbox{\hskip1pt}}}
\newcommand{\ho}[1]{$\, ^{#1}$}
\newcommand{\hoch}[1]{$\, ^{#1}$}
\newcommand{\bea}{\setlength\arraycolsep{2pt} \begin{eqnarray}}
\newcommand{\eea}{\end{eqnarray}}
\newcommand{\ra}{\rightarrow}
\newcommand{\lra}{\longrightarrow}
\newcommand{\Lra}{\Leftrightarrow}
\newcommand{\bp}{\tilde \beta^\prime}
\newcommand{\tr}{{\rm tr} }
\newcommand{\Tr}{{\rm Tr} }
\def\0{{\sst{(0)}}}
\def\1{{\sst{(1)}}}
\def\2{{\sst{(2)}}}
\def\3{{\sst{(3)}}}
\def\4{{\sst{(4)}}}
\def\5{{\sst{(5)}}}
\def\6{{\sst{(6)}}}
\def\7{{\sst{(7)}}}
\def\8{{\sst{(8)}}}
\def\m{{\sst{(m)}}}
\def\n{{\sst{(n)}}}
\def\cA{{{\cal A}}}
\def\cB{{{\cal B}}}
\def\cF{{{\cal F}}}
\def\cG{{{\cal G}}}
\def\cH{{{\cal H}}}
\def\tV{\widetilde V}
\def\tW{\widetilde W}
\def\tH{\widetilde H}
\def\tE{\widetilde E}
\def\tF{\widetilde F}
\def\tA{\widetilde A}
\def\im{{{\rm i}}}
\def\tY{{{\wtd Y}}}
\def\ep{{\epsilon}}
\def\vep{{\varepsilon}}
\def\bD{{{\bar D}}}
\def\R{{{\mathbb R}}}
\def\C{{{\mathbb C}}}
\def\H{{{\mathbb H}}}
\def\CP{{{\mathbb C}{\mathbb P}}}
\def\RP{{{\mathbb R}{\mathbb P}}}
\def\Z{{{\mathbb Z}}}
\def\bA{{{\mathbb A}}}
\def\bB{{{\mathbb B}}}
\def\bC{{{\mathbb C}}}
\def\bD{{{\mathbb D}}}
\def\bE{{{\mathbb E}}}
\def\bZ{{{\mathbb Z}}}
\def\Re{{{\frak{Re}}}}
\def\Im{{{\frak{Im}}}}
\def\cosec{{\,\hbox{cosec}\,}}
\def\Gm{{\Gamma_{\!\! -}}}
\def\Gp{{\Gamma_{\!\! +}}}
\def\stan{{standard }}
\def\nonstan{{supernumerary }}
\def\p{{\partial}}
\def\kdel#1{{\fft{\del}{\del#1}}}

\def\bog{{Bogomolny }}
\def\om{{\omega}}

\newcommand{\nnr}{\nonumber \\}
\newcommand{\pd}{\partial}
\newcommand{\ud}{\textrm{d}}
\newcommand{\dTH}{T^{\prime \, 0}_\textrm{H}}
\newcommand{\dOi}{\Omega^{\prime \, 0}_i}
\newcommand{\bx}{{\bf x}}

\begin{document}

\vspace{5mm}
\begin{center}
{\Large \bf Lifshitz black holes in the  Ho\v{r}ava-Lifshitz gravity
} \vspace{12mm}

{\large   Yun Soo Myung \footnote{e-mail
 address: ysmyung@inje.ac.kr}}
 \\
\vspace{10mm} {\em Institute of Basic Science and School of
Computer Aided Science \\ Inje University, Gimhae 621-749, Korea}
\end{center}

\begin{center}

\underline{Abstract}
\end{center}

We investigate the Lifshitz black holes from the Ho\v{r}ava-Lifshitz
gravity by comparing with the Lifshitz black hole from the 3D new
massive gravity.  We note that these solutions all have single
horizons. These black holes are very similar to each other when
studying their thermodynamics. It is shown that a second order phase
transition is unlikely possible to occur between $z=3,2$ Lifshitz
black holes and $z=1$ Ho\v{r}ava black hole.

\vspace{15pt}

\thispagestyle{empty}





\newpage
\section{Introduction}
Ho\v{r}ava has proposed a renormalizable theory of gravity at a
Lifshitz point~\cite{ho1,ho2},  which  may be regarded as a UV
complete candidate for general relativity. At short distances the
theory of  Ho\v{r}ava-Lifshitz (HL) gravity with a flow parameter
$\lambda$ describes interacting non-relativistic gravitons and is
supposed to be power counting renormalizable in (1+3) dimensions.
Recently, its black hole solutions have  been intensively
investigated
in~\cite{LMP,CCO1,CLS,CY,MK,CCO2,KS,Myung,park,Gho09,CL,PW,lkme,DCJ}.

There are two classes of Ho\v{r}ava-Lifshitz gravity in the
literature: {\it the projectable and nonprojectable theories} where
the former (latter) implies that the lapse function  depends on time
(time and space). A main issue of the Ho\v{r}ava-Lifshitz gravity is
still to answer to the question of whether it can
 accommodate the Ho\v{r}ava scalar $\psi$,
  in addition to two degrees of freedom (DOF) for a massless graviton.
To this end, we would like to mention  relevant works. The
authors~\cite{CNPS} have shown that in  the nonprojectable theories,
the Ho\v{r}ava scalar $\psi$ is related to a scalar degree of
freedom appeared in  the massless limit of a massive graviton.
Especially for the Hamiltonian approach to the HL gravity, the
authors~\cite{LP} did not consider the Hamiltonian
 constraint as a second class constraint, which leads to a  strange
 result that there are no DOF left when  imposing the
 constraints of the theory. Moreover, the authors~\cite{HKG} have claimed that
 there are no solution of the lapse function which  satisfies  the
 constraints. Unfortunately, it implies  a surprising conclusion that there is no
 evolution  at all  for any observable.   However, more recently, it
 was shown that the IR version of HL gravity ($\lambda
 R$-model)  is completely equivalent to the general relativity for any
 $\lambda$  when employing a consistent Hamiltonian formalism based on Dirac algorithm~\cite{BR}.

In the projectable theories, the authors~\cite{SVW,BPS} have argued
that $\psi$ is propagating around the Minkowski space but it has a
negative kinetic term, showing a ghost instability. In this case,
the Ho\v{r}ava scalar becomes ghost if the sound speed square
($c^2_\psi$) is positive. In order to make this scalar  healthy, the
sound speed square must be negative, but it is inevitably unstable.
Thus, one way to avoid this is to choose the case that the sound
speed square is close to zero, which implies the limit of $\lambda
\to 1$. However, in this limit, the cubic interactions are important
at very low energies, called the strong coupled problem~\cite{KA}.
This invalidates any linearized analysis and any predictability of
quantum gravity is lost due to unsuppressed loop corrections. This
casts serious doubts on the UV completeness of the theory.  The
authors~\cite{BPS2} tried to extend the theory to make a healthy HL
gravity, but there has been some debate as to whether this theory is
really healthy~\cite{PS,BPS3,KP}. The projectability condition from
condensed matter physics may not be appropriate for describing the
(quantum) gravity. Instead, if one does not impose the
projectability condition, the HL gravity leads to general relativity
without the  strong coupling problem in the IR limit.

On the other hand, the Lifshitz-type black
holes~\cite{CFT-4,L-1,AL-3,L-2,L-4,L-3,L-5} have received
considerable attentions since these may provide a model of
generalizing AdS/CFT correspondence to non-relativistic condensed
matter physics~\cite{CFT-1,CFT-2,CFT-3}. However, even though their
asymptotic spacetimes  are apparently simple as Lifshitz, the
problem of obtaining an analytic  solution seems to be a nontrivial
task.  Some examples include  a 4D topological black hole which is
asymptotically Lifshitz with the dynamical exponent
$z=2$~\cite{Mann}. An analytic black hole solution with $z=2$ that
asymptotes a planer Lifshitz spacetime was found in 4D
spacetimes~\cite{bm}, and numerical solutions were also explored in
~\cite{BBP,AL-2}.  The analytic examples of Lifshitz black holes in
higher dimensions were reported in~\cite{AGGBh}
Interestingly, the
$z=3$ Lifshitz black hole~\cite{z3} was derived from the new massive
gravity (NMG) in 3D spacetimes~\cite{bht}. It was claimed that there
is a subtle issue to define thermodynamic quantities of this  black
hole  because of negative mass and entropy~\cite{L-5}. However,  a
thermodynamic study for the $z=3$ Lifshitz black hole was performed
by using the 2D dilaton gravity approach~\cite{MKLdil}, showing that
its thermodynamics is rather simple and is consistently defined.
Also, a boundary stress-tensor approach to this black hole has
confirmed that the wrong (negative)-sign Einstein-Hilbert term
provides really a consistent thermodynamics of the $z=3$ Lifshitz
black hole~\cite{HT}.

Concerning a static spherically symmetric  solution, L\"u-Mei-Pope
(LMP) have obtained the black hole solution with
$\lambda$~\cite{LMP} and topological black holes were found in
\cite{CCO1}.    We remind the reader that these black hole solutions
were obtained from the Ho\v{r}ava-Lifshitz gravity without imposing
the projectability condition (nonprojectable theories). Within the
projectable theories, their black hole solutions are less
interesting~\cite{GPW}. Its thermodynamics was studied in
\cite{MK,CCO2}, but there remain unclear issues in obtaining the ADM
mass and the entropy because for $1/3\le \lambda \le 1/2$, the LMP
solution belongs to Lifshitz black holes with $2\le z\le 4$. In this
case, the entropy may  take a very unusual form as $S=A/4-
(\pi/\Lambda_W)\ln[A/4]$ with $A$ the area of horizon~\cite{CL}. It
was well known that many different kinds of black holes from string
theories have the Bekenstein-Hawking entropy of
$S_{BH}=A/4$~\cite{Ahar}.  Thus, one has to explain why a
logarithmic term $(-\pi/\Lambda_W)\ln[A/4]$ appears as a part of the
entropy of Lifshitz black hole in the HL gravity~\cite{myungent,CO}.
This term arises because one has used the first law of $dS=dm/T_H$
to derive the entropy, provided that the Hawking temperature  $T_H$
and the mass $m$ have been  known. Indeed, the mass $m$ was not
clearly defined by either the condition of the zero metric function
$f=0$~\cite{MK} or a Hamiltonian approach~\cite{CCO2}. Until now,
there is no definite way to calculate the Arnowitt-Deser-Misner
(ADM) mass $M_{ADM}$ for the Lifshitz black hole, if one insists
that the ADM mass should be evaluated at asymptotic Lifshitz.
 Hence, it would be better  to use the
Bekenstein-Hawking entropy to derive the horizon mass for Lifshitz
black holes when applying  the first law of $dM_{h}=T_H
dS_{BH}$~\cite{Myungadm,AL-2}.

In this work, we obtain  the horizon  mass of the Lifshitz  black
holes in the nonprojectable HL gravity. In deriving this mass, we
use the first law of thermodynamics and  the Bekenstein-Hawking
entropy. We investigate thermodynamics of $z=3,2$ Lifshitz black
holes in the nonprojectable  Ho\v{r}ava-Lifshitz gravity by
comparing with the $z=3$ Lifshitz black hole in the NMG. Finally, we
discuss a second order phase transition between the $z=3,2$ Lifshitz
black holes and the $z=1$ Ho\v{r}ava black hole.

 \section{HL gravity}
Introducing the ADM formalism where the metric is parameterized
\be ds_{ADM}^2= - N^2  dt^2 + g_{ij} \Big(dx^i - N^i dt\Big)
\Big(dx^j - N^j dt\Big)\,, \ee
the Einstein-Hilbert action can be expressed as
\be \label{Eins} S_{EH} = \fft{1}{16\pi G} \int d^4x \sqrt{g} N
\Big[K_{ij} K^{ij} - K^2 + R - 2\Lambda\Big], \ee
where $G$ is Newton's constant and extrinsic curvature $K_{ij}$
takes the form
\be K_{ij} = \fft{1}{2N} \Big(\dot g_{ij} - \nabla_i N_j -
\nabla_j N_i\Big)\,. \ee
Here, a dot denotes a derivative with respect to $t$. The ${\bf
Z}=3$ HL action of a non-relativistic  gravitational theory  is
given by~\cite{ho1} \be S_{HL}=\int dtd^3x \Big[{\cal L}_K + {\cal
L}_V\Big],  \label{action1} \ee where the kinetic Lagrangian is
given by \be {\cal L}_K = \frac{2}{\kappa^2}\sqrt{g}
N\Big(K_{ij}K^{ij}-\lambda K^2\Big). \ee The potential Lagrangian
is determined by the detailed balance condition as \bea {\cal
L}_V&=&
\sqrt{g}N\Bigg[\frac{\kappa^2\mu^2}{8(1-3\lambda)}\Big(\frac{1-4\lambda}{4}R^2
+\Lambda_WR-3\Lambda_W^2\Big)\nn \\
 &-&\frac{\kappa^2}{2w^4} \left(C_{ij}
-\frac{\mu w^2}{2}R_{ij}\right) \left(C^{ij} -\frac{\mu
w^2}{2}R^{ij}\right) \Bigg].\label{action2} \eea In the IR limit,
comparing (\ref{action1}) with (\ref{Eins}) of general relativity,
the speed of light, Newton's constant and the cosmological
constant are given by
\be c=\fft{\kappa^2\mu}{4}
\sqrt{\fft{\Lambda_W}{1-3\lambda}}\,,\qquad
G=\fft{\kappa^2}{32\pi\,c}\,,\qquad \Lambda_{\rm cc}=\ft32
\Lambda_W\,.\label{cg} \ee The equations of motion were derived in
\cite{cos1} and \cite{LMP}. In order to have a black hole
solution, it requires that  $\lambda>1/3$ and $\Lambda_W<0$
because the speed of light $c$ blows up at $\lambda=1/3$.

\section{3D Lifshitz black holes in the NMG}
The NMG action~\cite{bht} composed of the Einstein-Hilbert action
with a cosmological constant $\Lambda$ and higher order curvature
terms is given by
\begin{eqnarray}
\label{NMGAct}
 S^{(3)}_{NMG} &=& S^{(3)}_{EH}+S^{(3)}_{FH}, \\
\label{NMGAct2} S^{(3)}_{EH} &=& -\frac{1}{16\pi G_3} \int d^3x \sqrt{-\cal{G}}~ ({\cal R}-2\Lambda),\\
\label{NMGAct3} S^{(3)}_{HC} &=& \frac{1}{16\pi G_3u^2} \int d^3x
            \sqrt{-\cal{G}}~\left({\cal R}_{MN}{\cal R}^{MN}-\frac{3}{8}{\cal R}^2\right),
\end{eqnarray}
where $G_3$ is a 3D Newton constant and $u^2$ a parameter with mass
dimension 2. We note that the wrong-sign appears in the
Einstein-Hilbert term (EH) when comparing to the original
action~\cite{bht}. This means that we keep a relative ($-$) sign of
higher order curvature term fixed with respect to the EH.

The field equation is given by \be {\cal
R}_{MN}-\frac{1}{2}g_{MN}{\cal R}+\Lambda
g_{MN}-\frac{1}{2u^2}K_{MN}=0,\ee where
\begin{eqnarray}
  K_{MN}&=&2\square {\cal R}_{MN}-\frac{1}{2}\nabla_M \nabla_N {\cal R}-\frac{1}{2}\square{\cal R}g_{MN}\nonumber\\
        &+&4{\cal R}_{MNPQ}{\cal R}^{PQ} -\frac{3}{2}{\cal R}{\cal R}_{MN}-{\cal R}_{PQ}{\cal R}^{PQ}g_{MN}
         +\frac{3}{8}{\cal R}^2g_{MN}.
\end{eqnarray}
In order to have the  Lifshitz black hole solution with dynamical
exponent $z$, it is convenient to introduce dimensionless
parameters \be y=u^2~ \ell^2,~~w=\Lambda~ \ell^2, \ee where $y$
and $w$ are proposed to take \be
y=-\frac{z^2-3z+1}{2},~~w=-\frac{z^2+z+1}{2}. \ee For the $z=1$
non-rotating BTZ black hole, one has $y=\frac{1}{2}$ and
$w=-\frac{3}{2}$, while
 $y=-\frac{1}{2}$ and
$w=-\frac{13}{2}$ are chosen for the  $z=3$ Lifshitz black hole.
For $z=1$ and 3, the black hole solutions are given by \be
\label{2pdmetric}
  ds^2_{3D}=-x^{2z}F(r)dt^2
   +\frac{1}{x^2} H(r)dr^2+ r^2d\theta^2,
\end{equation}
where \be x=\frac{r}{\ell},~~F(r)=\frac{1}{H(r)}=1-\frac{{\cal
M}\ell^2}{r^2}=1-\frac{r_+^2}{r^2}.\ee We emphasize that  ${\cal M}$
is a mass parameter related to the horizon  mass $M$.  A naive
condition of $F(r)=0$ could not determine the horizon mass $M$ in
the $z=3$ Lifshitz black holes, as contrasts to $M=r_+^2/\ell^2$ for
the $z=1$ non-rotating BTZ black hole. The metric (\ref{2pdmetric})
implies that a curvature singularity appears at $r=0$ as is
shown~\cite{z3} \be {\cal R}=-\frac{26}{\ell^2}+\frac{8{\cal
M}}{r^2},~~{\cal R}_{MN}{\cal R}^{MN}=\frac{260}{\ell^4}-\frac{152
{\cal M}}{\ell^2r^2}+\frac{24 {\cal M}^2}{r^4},\ee
 and a single event horizon
is located at $r=r_+=\ell \sqrt{{\cal M}}$.
 For the $z=3$
Lifshitz black hole, its thermodynamic quantities of Hawking
temperature $T_H$, horizon mass $M \simeq {\cal M}^2$, heat capacity
$C=\frac{dM}{dT_H}$,  Bekenstein-Hawking entropy $S_{BH}$, and free
energy $F=M-T_HS_{BH}$ are given by~\cite{MKLdil} \be \label{thermo}
T_H=\frac{x_+^3}{2\pi \ell},~~M=\frac{x_+^4}{2},~~C=4\pi x_+,
~~S_{BH}=4\pi r_+,~~F=-\frac{3}{2}x_+^4 \ee with $x_+=r_+/\ell$. We
check that the above quantities satisfy the first law of
thermodynamics \be dM=T_H dS_{BH}. \ee

Now we are in a position to explain why we start with the wrong-sign
EH term in (\ref{NMGAct2}). When making a replacement of  $G_3 \to
-G_3$ to go back the original NMG~\cite{bht}, the temperature  is
the same, but mass and entropy are negative~\cite{L-5}.  Negative
mass and entropy are not permissible for black hole physicists and
thus, this problem  should be resolved. One way to resolve it  was
to replace  the Newton's constant $G_3$ by $-G_3$, leading to our
action (\ref{NMGAct}). It was shown that this replacement is indeed
necessary to regard the NMG as a unitary massive gravity~\cite{NO}.
The NMG is equivalent to the Fierz-Pauli massive gravity within the
linearized theory. In three dimensions, a massless graviton has no
DOF, while a massive graviton is a physically propagating mode with
two helicities. In constructing the NMG with higher order curvature
term, the important thing was that one can neglect a massless
graviton whatever its norm is positive or negative, in favor of a
massive graviton without ghost from (\ref{NMGAct3})~\cite{Oda}. In
this sense, our action (\ref{NMGAct}) is a reliable one  to derive
the correct thermodynamic quantities. Importantly,  we note that
{\it the higher order curvature term (\ref{NMGAct3})  of the NMG  is
not a perturbative correction to the Einstein-Hilbert action, but
the main term to obtain  the $z=3$ Lifshitz black hole.}  Recently,
a boundary stress-tensor approach to this black hole ~\cite{HT} has
confirmed that the wrong (negative)-sign Einstein-Hilbert term
provides really a thermodynamics of the $z=3$ Lifshitz black hole
which is consistent with (\ref{thermo}). Especially, a proper
definition for the mass of the $z=3$ Lifshitz black hole was
introduced.

For $z=1$ and 3, their quantities could be rewritten by the compact
forms \be \label{3dlbh}T^z_H = \frac{ x_+^z}{2\pi \ell},~~M^z =
\frac{ 2}{1+z}x_+^{z+1},~~ C=4 \pi x_+,~~S=4\pi r_+,~~
F^z=-\frac{2z}{1+z}x^{z+1}_+, \ee where the heat capacity and
Bekenstein-Hawking entropy remain unchanged.

Considering the free energy of $F^{z}$ and the coordinate matching
(see Fig. 1), there is a crossing point at $x_+=x_c=0.82$ where
$F^{z=1}(x_+)$ is equal to $F^{z=3}(x_+)$. This implies that for $0
\le x_+ \le x_c$, $F^{z=3} \ge F^{z=1}$, while for $x_+ \ge x_c$,
$F^{z=3} \le F^{z=1}$. That is, for $0 \le x_+ \le x_c$, the $z=1$
non-rotating BTZ  black hole is more favorable than the $z=3$
Lifshitz black hole, while for $ x_+\ge x_c$, the  $z=3$ Lifshitz
black hole is more favorable than the $z=1$ non-rotating BTZ black
hole. This may imply a second order phase transition between two
black holes~\cite{myungscalar}. However, we note that two black
holes have different asymptotes: Lifshitz and AdS$_3$ spacetimes.
Hence, the phase transition  occurs unlikely between two black
holes. In order to see it explicitly,  we use the other called the
temperature matching. When expressing  their free energy in terms of
their Hawking temperatures, one has \be F^{z=3}(T_H)=-\frac{3}{2} (
2\pi \ell T_H)^{\frac{4}{3}},~~F^{z=1}(T_H)=- ( 2\pi \ell T_H)^{2},
\ee which shows that for $0 \le T_H \le T_c$, $F^{z=3}(T_H) \le
F^{z=1}(T_H)$, while for $ T_H \ge T_c$, $F^{z=3}(T_H) \ge
F^{z=1}(T_H)$ with $T_c=\frac{1.82}{2\pi \ell}$. In Fig. 3, we use
the notation of $2\pi T_H \to T_H$ with $\ell=1$ for simplicity.
This means that the temperature matching provides the result which
is opposite to the coordinate matching.

Consequently, a  second order phase transition is unlikely possible
to occur between two black holes.
\begin{figure}[t!]
   \centering
   \includegraphics{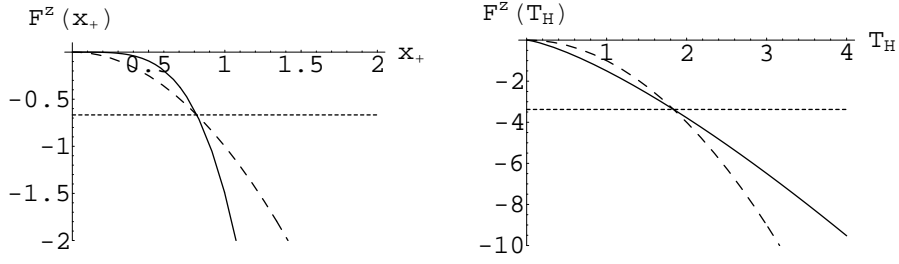}
\caption{Graphs of 3D free energy $F^{z}(x_+)$ and $F^z(T_H)$. Left:
two free energies (solid curve: $F^{z=3}$; dashed curve: $F^{z=1}$)
with coordinate matching cross at $x_+=x_c=0.82$. Right: two free
energies with temperature matching cross at  $ T_H=T_c=1.82$. Two
are opposite to each other around the crossing points.}
\label{fig.1}
\end{figure}

\section{4D Lifshitz black holes in the nonprojectable HL gravity}

A static spherically symmetric (SSS) solution to the nonprojectable
HL gravity was obtained by considering the line element
\begin{equation}
ds_{4D}^2 = -N(r)^2 dt^2 + \frac{dr^2}{f(r)} + r^2 \left ( d
\theta^2 + \sin^2 \theta d \phi^2 \right ) . \label{sss}
\end{equation}
Here the ``nonprojectable" notion is clear because the lapse
function $N$ depends on space coordinate $r$ only in the static
solution of the black hole.  For this purpose,  choosing $K_{ij}=0$
and $C_{ij}=0$, the Lagrangian reduces to the ${\bf Z}=2$ potential
Lagrangian as \be {\cal L}_V=\sqrt{g}N
\frac{\kappa^2\mu^2(-\Lambda_W)}{8(3\lambda-1)}
\Bigg[R-3\Lambda_W-\frac{4\lambda-1}{4\Lambda_W}R^2
+\frac{(3\lambda-1)R^2_{ij}}{\Lambda_W} \Bigg].\ee  Substituting the
metric ansatz (\ref{sss}) into ${\cal L}_V$, one has the reduced
Lagrangian
\begin{eqnarray}
\label{react} {\cal L}^{SSS}_V=\frac{ \kappa^2\mu^2 (-\Lambda_W) N
}{8(1-3\lambda)\sqrt{f}}\Bigg[&-&2(1 - f - r f')+3 \Lambda_Wr^2  \\
\nn &-& \frac{(2\lambda-1)(f-1)^2}{\Lambda_W r^2}
-\frac{\lambda-1}{2\Lambda_W} f'^2 + \frac{2\lambda (f-1)}{\Lambda_W
r}f'\Bigg],
\end{eqnarray}
where $^\prime$ denotes the differentiation with respect to $r$. We
note that $ {\cal L}^{SSS}_V$ is not appropriately defined at
$\lambda=1/3$ because it blows up at this point. Hereafter, we
exclude $\lambda=1/3$ from our consideration.
 The
L\"{u}-Mei-Pope (LMP) solution for the HL gravity is  given by \be
f(x) = 1 + x^2 - \alpha x^{p_\pm(\lambda)},~~ N(x) =
x^{q_\pm(\lambda)} \sqrt{f(x)}, \label{bh-sol} \ee where $\alpha$ is
an integration constant related to the horizon  mass of the black
hole and \be x = \sqrt{-\Lambda_W} r,~ p_\pm(\lambda) = \frac{2
\lambda \pm \sqrt{6 \lambda -2}}{\lambda -1},~~q_\pm(\lambda) =-
\frac{1 + 3 \lambda \pm 2 \sqrt{6 \lambda -2}}{\lambda -1}.\ee
 In this work, we choose
$p_-(\lambda)=p$ and $q_-(\lambda)=q$ and thus, $2p+q=1$. In this
case, it was shown that for $p <2$, the LMP solution is singular at
$r=0$~\cite{CW} as \be R=6\Lambda_W+ \frac{2\alpha(1+p)x^p}{r^2}.
\ee Here we note that for $p=-1(\lambda=1/3)$, 3D Ricci scalar $R$
is regular at $r=0$, which implies  that the $\lambda=1/3$ does not
correspond to a Lifshitz black hole.
 Its
extremal black hole with $f(x_e) =0$ and $ f'(x_e) = 0$ are
located as \be \label{extrecon} x_e=0,~{\rm for}~\frac{1}{3}<
\lambda \le \frac{1}{2};~~
x_e=\sqrt{\frac{p}{2-p}}=\sqrt{\frac{2\lambda-\sqrt{6\lambda-2}}{-2+\sqrt{6\lambda-2}}},~{\rm
for}~\lambda
> \frac{1}{2}. \ee
However, assuming the near-horizon geometry of AdS$_2\times S^2$,
the radius $v_2$ of $S^2$ is negative for $1/3< \lambda \le 1/2$,
which means that the near-horizon geometry of extremal black hole is
ill-defined and the corresponding Bekenstein-Hawking entropy is
zero~\cite{lkme}. For $\lambda>1/2$, the  near-horizon geometry of
extremal black holes are AdS$_2 \times S^2$ with different radii,
depending on the  HL gravity. This shows clearly  that for $1/3 <
\lambda \le 1/2$, the horizon at $x_e=0$ is a single horizon but not
a degenerate horizon. Actually, these correspond to  the Lifshitz
black holes with $2\le z < 4$~\cite{MK}.

 In order to understand this
branch  of the LMP black holes fully, it is necessary to introduce
4D Lifshitz black holes.  Their line element  takes the form with
dynamical exponent $z$~\cite{L-1,AL-3,Mann,z3} \be \label{lifsh}
ds^2_{\rm Lif}=- x^{2z}
F(r)dt^2+\frac{1}{x^2}H(r)dr^2+r^2d\Omega^2_{2}, \ee where $F(r)$
and $H(r)$ are functions of a radial coordinate $r$ with \be \lim_{r
\to \infty}F(r)=\lim_{r \to \infty}H(r)=1. \ee Comparing the LMP
black holes (\ref{sss}) with the Lifshitz black holes (\ref{lifsh})
leads to  the correspondence \be
N^2=\tilde{N}^2f=x^{2(q+1)}\frac{f}{x^2} \to x^{2z}F(r),~~f \to
\frac{x^2}{H(r)} \ee which imply the two relations \be z=q+1,~~F(r)=
\frac{1}{H(r)}=\frac{f}{x^2}. \ee This indicates how  the 4D
Lifshitz black holes originate  from a non-relativistic theory of
the HL gravity. We note that the $z=3,2$ Lifshitz black holes are
obtained from either ${\bf Z}$=2 or 3 HL action, which means that a
relevant quantity to determine the exponent $z$ is not the dynamical
scaling dimension ${\bf Z}$ but the flow parameter $\lambda$ through
$q$ ($\lambda=0.36 \to z=3,~\lambda=1/2 \to z=2$). As far as the LMP
solution is concerned, there is no distinction between ${\bf Z}=2$
and ${\bf Z}=3$ HL gravities because of $C_{ij}=0$. It seems that
this is a feature of the HL gravity, compared with the NMG.

\section{Thermodynamics of 4D Lifshitz black holes}
In order to explore  properties of the 4D Lifshitz back holes, let
us first study the Hawking temperature  because it is derived from
the the surface gravity $\kappa$ defined at the horizon and thus,
is really independent of the mass parameter $\alpha$. The Hawking
temperature is defined by
\be
\label{temp}T^z_H(x_+)=\frac{\kappa}{2\pi}=\frac{\sqrt{-\Lambda_W}}{8\pi}\Big[(z+2)x_+^{z}+(z-2)
~x^{z-2}_+\Big], \ee where we use the relations of $q=z+1$ and
$p=\frac{2-z}{2}$ in deriving the last expression. As is shown in
Fig. 1, for $2 \le z <4$, it is a monotonically increasing
function like $T\simeq x_+^{z}$ for large $x_+$.
\begin{figure}[t!]
   \centering
   \includegraphics{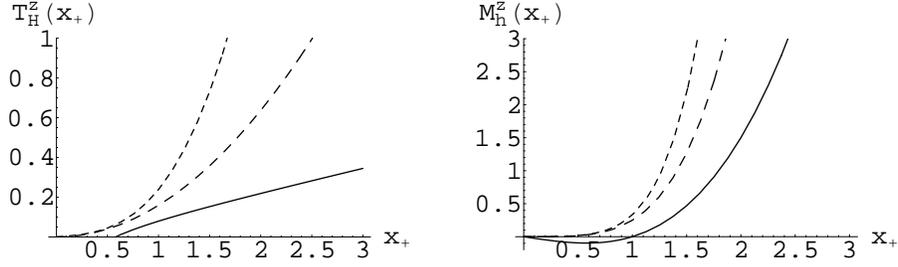}
\caption{Graphs of Hawking temperature $T^{z}_H(x_+)$ and horizon
mass $M^{z}_h(x_+)$ with $\Lambda_W=-1(x_+=r_+)$. Left: two dashed
curves represent the Hawking temperatures for $z=3,2$ from top to
bottom and the solid curve denotes the $z=1 $ Hawking temperature
which is positive for $x_+>1/\sqrt{3}$. Right: two dashed curves
represent the horizon masses for $z=3,2$ from top to bottom and the
solid curve denotes the $z=1$ horizon mass which is positive only
for $x_+>1$.} \label{fig.2}
\end{figure}
In order to find the horizon mass, we may  use the first law of
thermodynamics \be dM_h=T_HdS_{BH}, \ee where the Bekenstein-Hawking
entropy\footnote{ The other entropy of $S=\pi
r_+^2-\frac{\pi}{\Lambda_W}\ln[\pi r_+^2]$ could be obtained from
the first law of $dS=\frac{dm}{T_H}$ provided the mass of $m \simeq
\alpha^2 $ and the temperature (\ref{temp}) were
known~\cite{CCO1,CCO2,myungent,CO}. However, it is hard to accept
this entropy because one could not have a logarithmic term unless
either thermal correction or quantum correction is considered. In
the HL gravity, {\it higher order curvature term} plays an essential
role to make the Lifshitz black hole. This is not a correction to
the wrong-sign EH term. Hence, the Bekenstein-Hawking entropy is
more natural to derive the horizon mass.} satisfies the area-law of
\be \label{bhent} S_{BH}=\pi r_+^2.\ee
 Then,
the horizon mass is obtained as \be  \label{admmass}
M^z_h(x_+)=\int^{x_+}_0 T_H
dS_{BH}=\frac{x_+^{z}}{4\sqrt{-\Lambda_W}}\Big[x^2_+
+\frac{z-2}{z}\Big]. \ee For $z=3,2$, the horizon masses are given
by, respectively,  \be
M^{z=3}_h=\frac{x_+^3}{4\sqrt{-\Lambda_W}}\Big[x_+^2+\frac{1}{3}\Big],~~
M^{z=2}_h=\frac{x_+^{4}}{4\sqrt{-\Lambda_W}}.\ee We could not
determine the $z=0(\lambda=3)$ horizon  mass because the last term
in (\ref{admmass}) blows up at $z=0$.

As is depicted in Fig. 2, the horizon mass is a monotonically
increasing function like $M^z_h \simeq x_+^{z+2}$ for $2\le z <4$
and  large $x_+$.   We note that the horizon mass is different from
the Komar charge defined as \be M^z_H=2T_H
S_{BH}=\frac{x_+^{z}}{4\sqrt{-\Lambda_W}}\Big[(z+2)x^2_++(z-2)\Big].\ee
\begin{figure}[t!]
   \centering
   \includegraphics{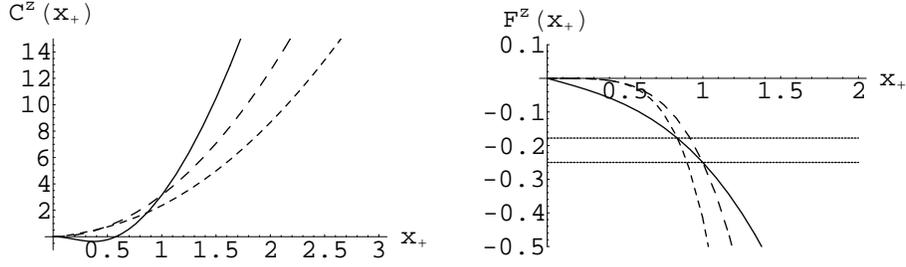}
\caption{Graphs of heat capacity $C^z(x_+)$ and free energy
$F^z(x_+)$ with $\Lambda_W=-1(x_+=r_+$). Left: two dashed curves
represent the heat capacity for $z=2,3$ from top to bottom and the
solid curve denotes the $z=1$ heat capacity which is negative for
$x_+<1/\sqrt{3}$. Right: two dashed curves of $z=2,3$ from top to
bottom cross the solid curve of the $z=1$ free energy at
$x_+=x_c=1,0.84$, respectively which are greater than
$x_+=x_e=1/\sqrt{3}$.} \label{fig.3}
\end{figure}
The heat capacity is an important quantity to test the thermal
stability: for $C>0$ the system is stable against thermal
perturbations, while for $C<0$, it is unstable thermodynamically.
This is defined by
\begin{eqnarray}
C^z(x_+)&=&\frac{dM^z_h}{dT^z_H}\nn \\
&=&2S_{BH}\Bigg[\frac{(z+2)x_+^{z-1}+(z-2)x_+^{z-3}}{z(z+2)x_+^{z-1}+(z-2)^2
x_+^{z-3} }\Bigg].
\end{eqnarray}
We observe from Fig. 3 that the heat capacity is a monotonically
increasing function like $C \simeq S_{BH} \simeq x_+^2$ for $ 2 \le
z <4$ and large $x_+$, which means that all Lifshitz black holes are
thermodynamically  stable because of $C\ge 0$. The free energy is
necessary to study a phase transition to other configuration. The
free energy is defined by \be
F=M^z_h-T^z_{H}S_{BH}=-\frac{x_+^{z}}{8\sqrt{-\Lambda_W}}\Big[zx^2_++\frac{(z-2)^2}{z}\Big].\ee
As is shown  in Fig. 3, all free energies are always negative.

 On the other hand, the case of
$z=1$ leads to different thermodynamic quantities as \be
\label{z=1lbh} T^{z=1}_H= \frac{3x_+^2-1}{8\pi
r_+},~~M^{z=1}_h=\frac{r_+}{4}\Big[x^2_+-1\Big],~~C^{z=1}=2 \pi
r_+^2\Bigg[\frac{3x_+^2-1}{3x^2_++1}\Bigg],~~F^{z=1}=-\frac{r_+(x^2_++1)}{8}\ee
whose asymptotic forms are given by \be T^{z=1}_H \simeq
x_+,~~M^{z=1}_h\simeq x_+^3,~~C^{z=1}\simeq x_+^2,~~F^{z=1}\simeq
-x^{3}_+. \ee This means that $\lambda=1$ LMP black hole is just the
$z=1$ Ho\v{r}ava  black hole. We call this the $z=1$ Ho\v{r}ava
black hole because its near-horizon geometry of the extremal black
hole is AdS$_2 \times S^2$ and its asymptote is AdS$_4$
spacetimes~\cite{LMP}, implying that it is basically different from
the $z=3,2$ Lifshitz  black holes.  At the extremal point of
$x_+=x_e=1/\sqrt{3}$, we have thermodynamic properties of
$T^{z=1}_H(x_e)=C^{z=1}(x_e)=0$ and
$(dM^{z=1}_h/dx_+)|_{x_+=x_e}=0$. If one uses the entropy of
$S=A/4-(\pi/\Lambda_W)\ln[A/4]$ to derive quasinormal modes of this
black hole~\cite{Maj}, the area spacing is not equidistant. Here, we
have equidistant area spacing because of the Bekenstein-Hawking
entropy.

All solid curves in Figs. 2 and 3 represent thermodynamic quantities
for the $z=1$ Ho\v{r}ava black hole.

For large Lifshitz black holes with $x_+ \gg 1$, their forms of
thermodynamic quantities are given by \be \label{4dlbh} T^z_H
\simeq x^z_+,~~M^z_h\simeq x_+^{z+2},~~C\simeq x_+^2,~~F^z \simeq
x_+^{z+2}, \ee while the Bekenstein-Hawking entropy (\ref{bhent})
remains unchanged.

Finally, comparing (\ref{4dlbh}) with (\ref{3dlbh}), their forms are
very similar to each other, but the difference is  exponents of
$M_h,~C,~S_{BH},$ and $F$ arisen from  the dimensionality.

\section{Phase transitions}

We discuss a possible phase transition by considering the coordinate
matching.  We note that the free energy of $F^{z=1}$ in
Eq.(\ref{z=1lbh}) is available only for $x_+>x_e=1/\sqrt{3}=0.58$
because one could not define the positive temperature for
$x_+<x_e=1/\sqrt{3}$. In this case, as is shown Fig. 3, there are
two crossing points at $x_+=x_c=0.84,1$ where
$F^{z=1}(x_+)=F^{z=3,2}(x_+)$. This implies that for $x_e \le x_+
\le x_c$, $F^{z=3,2} \ge F^{z=1}$, while for $x_+ \ge x_c$,
$F^{z=3,2} \le F^{z=1}$. In other words, for $x_e \le x_+ \le x_c$,
the $z=1$ black hole is more favorable than $z=3,2$ Lifshitz black
holes, while for $ x_+\ge x_c$, the $z=3,2$ Lifshitz black holes are
more favorable than the $z=1$ Ho\v{r}ava black hole. This may imply
a second order phase transition between Lifshitz black holes and
$z=1$ Ho\v{r}ava black hole.  It seems that this phase transition is
 related  to the second order phase transition between black hole
with scalar hair (Martinez-Troncoso-Zanelli black hole~\cite{MTZ})
and topological black hole with $k=-1$~\cite{KMPS,myungscalar},
which have the same AdS$_4$ asymptotes. However, the coordinate
matching is not an appropriate choice for investigating a second
order phase transition. The appropriate one is the temperature
matching. Unfortunately, we could not make a second order phase
transition between $z=3,2$ Lifshitz black holes and $z=1$ Ho\v{r}ava
black hole when using the temperature matching because they have
different asymptotes. It is clear from  Fig. 2 that  there is no
crossing point between $T^{z=3,2}_H(x_+)$ and $T^{z=1}_H(x_+)$ to
make the temperature matching.

\section{Discussions}

First of all, we mention that  4D Lifshitz black holes came from the
nonprojectable HL gravity, not the projectable theory. The
projectability condition from condensed matter physics may not be
appropriate for describing the (quantum) gravity. Especially for
static spherically symmetric solutions, the non-projectability
condition is necessary to obtain 4D Lifshitz black holes.

It is important to note that 3D Lifshitz and 4D Lifshitz black holes
are obtained only when including  higher order curvature terms, even
though the former action  is Lorentz-invariant combination with
wrong-sign Einstein-Hilbert action and the latter is
Lorentz-violating combination. This mean that these Lifshitz black
holes have purely gravity origin without introducing  any matter
field. We emphasize that {\it the role of higher order curvature
terms is essential for obtaining these black holes,} where these
terms are not considered simply as perturbative corrections to the
Einstein-Hilbert action.  If these terms are absent, one finds the
$z=1$ non-rotating BTZ black hole from the 3D Einstein gravity and
the Schwarzschild-AdS black hole from the 4D Einstein gravity.

These Lifshitz black holes  have much similarities. Their horizons
are non-degenerate. Their thermodynamic quantities are derived from
the Hawking temperature and the Bekenstein-Hawking entropy when
using the first law of thermodynamics. In this approach, we cannot
derive the horizon mass unless the Bekenstein-Hawking entropy is
used. We may insist  that  the area-law of the black hole entropy
and the first law of thermodynamics are valid for  Lifshitz black
holes~\cite{L-5,HT}. In this direction,  the dynamical evolution of
a massless scalar perturbation was investigated in the 4D Lifshitz
black hole spacetimes with the dynamical exponent
$z=4(\lambda={1/3})$, $z=2(\lambda={1/2})$ and $z=0(\lambda=3)$,
respectively~\cite{DCJ}. It has shown that scalar perturbations
decay without any oscillation in which the decay rate may imprint
thermodynamic quantities of the 4D Lifshitz black holes. These
purely damped modes are different from those in the (small)
Schwarzschild-AdS black hole but are similar to those in the 3D
charged black hole~\cite{MKPcd,Fer}, supporting that thermodynamic
nature of 4D Lifshitz black holes is simple. However, we are still
lacking for deriving the ADM mass of the Lifshitz-type black holes
using the Hamiltonian formalism because we do not know Lifshitz
asymptotes precisely.

These Lifshitz black hole spacetimes have curvature singularities at
$r=0$, and their asymptotes are Lifshitz. Also there is no phase
transition between Lifshitz black holes and $z=1$ black hole,
although there are crossing points between two free energies when
considering the coordinate matching. This is mainly  because their
asymptotes are different: Lifshitz and anti-de Sitter spacetimes.

Consequently, we have understood thermodynamics of Lifshitz black
holes  and have discussed possible second order phase transitions
between Lifshitz black holes and $z=1$ black holes.

\section*{Acknowledgement}
The author thanks Hyung Won Lee for helpful discussions.
 This work was in part  supported  by Basic Science
Research Program through the National Research  Foundation (NRF)
of Korea funded by the Ministry of Education, Science and
Technology (2009-0086861).

\end{document}